\documentclass[twocolumn,letterpaper]{IEEEAerospaceCLS}  %

\usepackage[]{graphicx}    %
\usepackage{amsmath, bm, physics}
\usepackage{float}
\usepackage{hyperref}
\newcommand{\ignore}[1]{}  %
\begin{document}
\title{Physics-Informed Neural Networks for Satellite State Estimation}

\author{%
    Jacob Varey\\
    MIT Lincoln Laboratory\\
    jacob.varey@ll.mit.edu
\and
    Jessica D. Ruprecht\\
    MIT Lincoln Laboratory\\
    jessica.ruprecht@ll.mit.edu
\and
    Michael Tierney\\
    MIT Lincoln Laboratory\\
    michael.kotson@ll.mit.edu
\and
    Ryan Sullenberger\\
    MIT Lincoln Laboratory\\
    ryan.sullenberger@ll.mit.edu
\thanks{\footnotesize 979-8-3503-0462-6/24/$\$31.00$ \copyright2024 IEEE}              %
\thanks{DISTRIBUTION STATEMENT A. Approved for public release. Distribution is unlimited.}
\thanks{Research was sponsored by the United States Air Force Research Laboratory and the Department of the Air Force Artificial Intelligence Accelerator and was accomplished under Cooperative Agreement Number FA8750-19-2-1000. The views and conclusions contained in this document are those of the authors and should not be interpreted as representing the official policies, either expressed or implied, of the Department of the Air Force or the U.S. Government. The U.S. Government is authorized to reproduce and distribute reprints for Government purposes notwithstanding any copyright notation herein.}
}

\maketitle

\thispagestyle{plain}
\pagestyle{plain}

\maketitle

\thispagestyle{plain}
\pagestyle{plain}

\begin{abstract}
The Space Domain Awareness (SDA) community routinely tracks satellites in orbit by fitting an orbital state to observations made by the Space Surveillance Network (SSN). In order to fit such orbits, an accurate model of the forces that are acting on the satellite is required. Over the past several decades, high\-quality, physics\-based models have been developed for satellite state estimation and propagation. These models have widely varying degrees of fidelity: some only account for two\-body Keplerian motion, while others consider highly accurate Earth gravity models, atmospheric drag, solar radiation pressure (SRP), perturbations from the Sun, Moon, and other celestial bodies, etc. These models are exceedingly good at estimating and propagating orbital states for non\-maneuvering satellites; however, there are several classes of anomalous accelerations that a satellite might experience which are not well\-modeled. For example, satellites using low\-thrust electric propulsion to modify their orbit, or pieces of debris which have extremely High Area\-to\-Mass Ratios (HAMR) that experience SRP effects beyond what is accounted for in existing models. Physics\-Informed Neural Networks (PINNs) are a valuable tool for these classes of satellites as they combine physics models with Deep Neural Networks (DNNs), which are highly expressive and versatile function approximators. By combining a physics model with a DNN, the machine learning model need not learn the fundamental physics of astrodynamics, which results in more efficient and effective utilization of machine learning resources to solve for only the unmodeled dynamics. This paper details the application of PINNs to estimate the orbital state and a continuous, low\-amplitude anomalous acceleration profile for satellites. Angles\-only observation data is simulated for satellites near Geosynchronous Orbit (GEO). This simulation first propagates the satellite in time using a physics model coupled with an arbitrary acceleration profile, and then simulates realistic, angles\-only observations of the satellite as would be measured by a ground\-based optical telescope. This arbitrary acceleration profile could represent a low\-thrust orbit maneuver or a difficult\-to\-model SRP effect on a HAMR object, and is used for generating truth data. The PINN is trained to learn the unknown acceleration by minimizing the mean square error of the observations. We evaluate the performance of pure physics models with PINNs in terms of their observation residuals and their propagation accuracy beyond the fit span of the observations. For a two\-day simulation of a GEO satellite using an unmodeled acceleration profile on the order of $\bm{10^{-8}\: km/s^2}$, the best\-fit physics model resulted in observation residuals with a root\-mean\-square error of 123 arcsec, while the best\-fit PINN had an error of 1.00 arcsec, comparable to the measurement noise. Similarly, after propagating the best\-fit physics model for five days beyond the fit span of the observations, the propagated position of the satellite using the physics\-only model was wrong by 3860 km, compared to the PINN which had an error of only 164 km. 
\end{abstract}

\tableofcontents

\section{Introduction} \label{section:intro}
For decades the 18th Space Defense Squadron (18th SDS), a component of the United States Department of Defense (DoD), has maintained a catalog of space objects \cite{18thSDS}. The Department of Commerce (DoC) has recently been tasked with providing Space Traffic Management (STM) services to commercial space operators \cite{space_policy_directive3}. While satellite owners and operators usually have exquisite telemetry from their satellites, these government organizations generally do not have access to that information. Instead, the Space Domain Awareness (SDA) community relies on observations of each Resident Space Object (RSO) from the Space Surveillance Network (SSN)--- a suite of optical and radar sensors dedicated to space surveillance \cite{shepherd2006space}. The 18th SDS maintains a satellite's orbit by fitting an orbital state--- either a Two Line Element (TLE) set or position and velocity vectors--- to observations correlated with the object. That orbital state is used to predict where the satellite will be in the future so that new observations can be correlated with the satellite. 

For the first several decades after sending the first satellites into Earth orbit, the challenge of maintaining knowledge of where each satellite was in space was driven by our lack of understanding of the natural forces acting on the satellite. More recently, high-fidelity physical models have been developed which can be used to quickly and accurately predict where a satellite will be at any time, assuming the satellite is not actively using thrusters to change its orbit. For example, the Special General Perturbations 4 (SGP4) propagator, which is used to predict a satellite's position and velocity from a TLE, models Earth gravity, atmospheric drag, and gravitational perturbations due to the Sun and Moon, but does not account for satellite thrust \cite{vallado2006revisiting}. Since the early days of human utilization of space, the assumption that a satellite is not actively thrusting has been violated exceedingly often. The vast majority of space missions today maintain a precise orbit using some form of thrust. In particular, electric propulsion (ion propulsion, Hall effect thrusters, pulsed plasma thrusters, etc.) are often utilized due to their increased fuel efficiency; the mass savings from requiring less fuel translates directly into more affordable satellite systems.

\begin{figure}
    \centering
    \includegraphics[width=\linewidth]{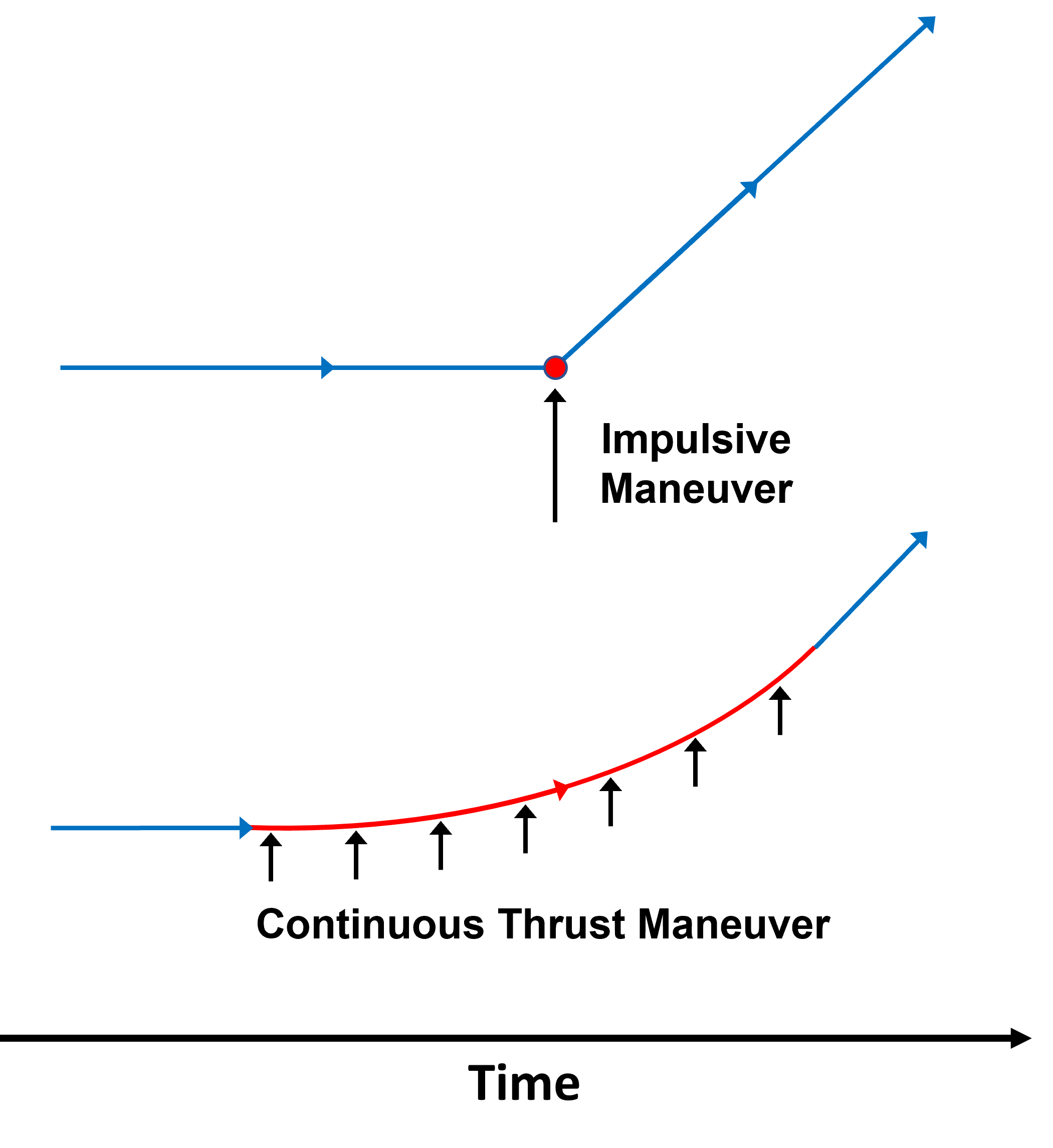}
    \caption{A diagram illustrating orbital state changes using impulsive and continuous thrust maneuvers. The regions shown by straight blue lines represent times in which the satellite is not thrusting, and is thus on a constant-energy orbit that is well-modeled by current propagators. The regions shown in red are those in which the satellite is thrusting and the assumption of the propagators is violated.}
    \label{fig:maneuver_cartoon}
\end{figure}

One important difference between electric propulsion systems and chemical thrusters is that chemical thrusters can achieve a large acceleration in a matter of seconds, while electric propulsion systems apply low thrust continuously for minutes to days. This difference is depicted in Figure \ref{fig:maneuver_cartoon}. Any amount of thrust violates the assumptions made by the physical model used to propagate the satellite's state from one time to another; however, the satellite's new orbit can be fit from post-maneuver observations. For satellites that use chemical thrusters to apply a large acceleration over a short period of time, the maneuvers are nearly instantaneous, or impulsive, and the satellite quickly changes from one orbit to another. As long as there are a sufficient number of new observations of the satellite after the maneuver, the satellite's new orbit can be determined because it is once again not thrusting in violation of the assumptions of the physical model. Because the change in orbital state happens rapidly for impulsive maneuvers, there is a window of opportunity to task sensors, correlate new observations of the satellite, and estimate a new, post-maneuver state; this is difficult for satellites that are continuously thrusting for long periods of time. Because continuously thrusting satellites spend long periods of time violating the assumptions of the physical model, by the time the maneuver has finished and a new orbit can be fit, the satellite may be be very far from where it was when the maneuver started, making the process of tasking and positively correlating new observations to the satellite's previous orbital state difficult.

Satellites employing electric propulsion for slow orbital maneuvers are difficult to maintain using pure physics models such as those built into SGP4. Because these models do not account for the RSO actively thrusting, the best-fit orbital state does not fit the observation data well and cannot be used to extrapolate far into the future. In practice, these satellite states are maintained with significant manual intervention. The process is the same as for any satellite: a new state is fit to a span of observations, which is then used to task sensors and correlate new observations. But because the state is a poor fit to the data, and because it does not extrapolate well into the future, this process must be repeated often to keep up with the ever-changing orbital state. If new observations of the satellite are not received quickly (e.g. due to a weather outage at an optical sensor site), it is possible that the RSO will be lost entirely. Sometimes new observations of the satellite are obtained, but they fail to correlate with the satellite's stale orbital state, and the observations will instead become an uncorrelated track (UCT). This challenge can be mitigated by estimating the satellite thrust as part of the orbit fitting process, making it possible to recover maneuvering objects despite a gap in observations.

Electric propulsion systems are now used for both small, station-keeping maneuvers, as well as large, orbital transfer maneuvers. Eutelsat 115 West B is an example of a satellite which used electric propulsion to raise its orbit from LEO to GEO over the course of nearly eight months using four Xenon Ion Propulsion System (XIPS) engines and now uses the same propulsion system for station-keeping in GEO \cite{thomas2016comparison}. Maintaining an orbital state on such a satellite during the orbital transfer period is more difficult than during the station-keeping period due to the magnitude and duration of the acceleration exerted by the thrusters. The goal of this work is to enhance the predictive power of an orbital state such that it can be used to task sensors and correlate new observations further into the future as compared to current models. That is, it is not necessary for the model to accurately predict where the satellite is for the entire eight month orbital transfer. Instead, the model needs to be able to accurately predict where the satellite will be a small number of days into the future-- long enough to task sensors to collect new observations, or to positively identify serendipitous observations to the satellite.

Let the \textit{thrust profile} be the magnitude and direction of the satellite's thrust vector as a function of time. The goal of this paper is to demonstrate the utility of Physics-Informed Neural Networks (PINNs) for describing the totality of the dynamics for a thrusting satellite. 

A PINN in the context of this paper refers to a model that combines physics and machine learning (ML), often a Deep Neural Network (DNN). This area of research is referred to as \textit{Knowledge-Informed Machine Learning} \cite{kiml2023}. Physical models are mathematical equations which have been developed and refined over time to accurately describe the physical world. In contrast, DNNs are highly expressive and versatile function approximators. By utilizing a PINN, the ML model need not relearn the fundamental physics, and can focus on learning just the deviations from the physical model present in the data, which is a more efficient use of ML resources. PINNs are an attractive solution for modeling the dynamics of a thrusting satellite because they combine the known physics (i.e. astrodynamics), which describes the natural forces acting on the satellite, with an ML model for describing the satellite thrust profile.

\subsection{Orbit State Fitting and Propagation}
Orbit fitting is the process of tuning the state parameters (i.e. the three-dimensional position and velocity vectors) to minimize the difference between the predicted and actual observations of the satellite. Predicted observations are generated by propagating the initial state of the satellite to the set of time epochs at which observations of the satellite have been made and using geometry to compute the expected observation based on the sensor site location and the sensing modality. For satellite tracking, the most common sensing modalities are active RF and passive optics. Radar sensors usually produce precise range and range-rate information, as well as less-precise angle and angle-rate information. Conversely, passive optical telescopes usually produce precise angles-only observations, but cannot determine range or range-rate. This paper focuses on orbit fitting using simulated observations from a single optical sensor observing a satellite in Geosynchronous (GEO) orbit, though the methodology is applicable across sensing modalities and orbit regimes.

Orbit propagation broadly falls into two types of techniques: semi-analytic methods and numerical integration methods. The former predict how a satellite's state will evolve based on in-depth analysis of the perturbing forces acting on the satellite and how those forces affect the satellite's state. Numerical integration methods differ in that they only estimate the total force acting upon the satellite as a function of the satellite state and time, and then numerically integrate the acceleration using an ordinary differential equation (ODE) solver to determine position and velocity as a function of time. For the task of propagating a satellite state in time, let the state vector be the concatenation of the three-dimensional position and velocity vectors of the satellite. The derivative of the state, then, is simply the velocity and acceleration vectors:

\begin{align}
    \bm{X} &= \begin{bmatrix}
    \bm{r} \\ 
    \bm{v}
    \end{bmatrix} 
    \label{eqn:state} \\
    \dot{\bm{X}} &= \begin{bmatrix}
        \dot{\bm{r}} \\
        \dot{\bm{v}}
    \end{bmatrix}
    = \begin{bmatrix}
        \bm{v} \\
        \bm{a}
    \end{bmatrix}
    \label{eqn:state_deriv}
\end{align}

These equations define an ODE with initial conditions (ICs):
\begin{equation} \label{eqn:initial_conditions}
    \bm{X_0} = \begin{bmatrix}
        \bm{r} \\
        \bm{v}
        \end{bmatrix}_0
\end{equation}

The satellite acceleration can be described with Cowell's formulation \cite{vallado2001fundamentals}:

\begin{equation} \label{eqn:acceleration_components}
    \bm{a} = -\frac{GM}{r^3}\bm{r} + \bm{a_P} + \bm{a_T}
\end{equation}

where the first term is the two-body equation of motion \cite{montenbruck_gill}, $G$ is the gravitational constant, $M$ is the mass of the primary body, $\bm{a_P}$ is the acceleration due to natural perturbing forces (e.g. non-uniformities in Earth's gravitational field, SRP, Sun and Moon gravity, etc.), and $\bm{a_T}$ is any acceleration the satellite is itself exerting via its thrusters. Precise physical models exist for the first two terms of equation \ref{eqn:acceleration_components}. The thrust component is often ignored, but modeling the satellite's thrust profile is required to accurately integrate the dynamics for a maneuvering object. The goal of this work is to demonstrate a model architecture that is flexible enough to learn a satellite's thrust profile directly from observation data. 

\section{Simulating Observation Data} \label{section:sim_data}
The first challenge in developing a model that accurately estimates the dynamics of a continuously thrusting satellite is acquiring observation data of such a satellite along with true ephemeris information describing where the satellite is as a function of time well into the future. Observation data is needed because that is all that is typically available when fitting an orbit, and the ephemeris data is needed for evaluating how well a model is able to extrapolate into the future, which is a key measure of performance. Due to the need for truth data, the demonstration of the technique described in this paper is performed using simulated data. 

\begin{figure}
    \centering
    \includegraphics[width=\linewidth]{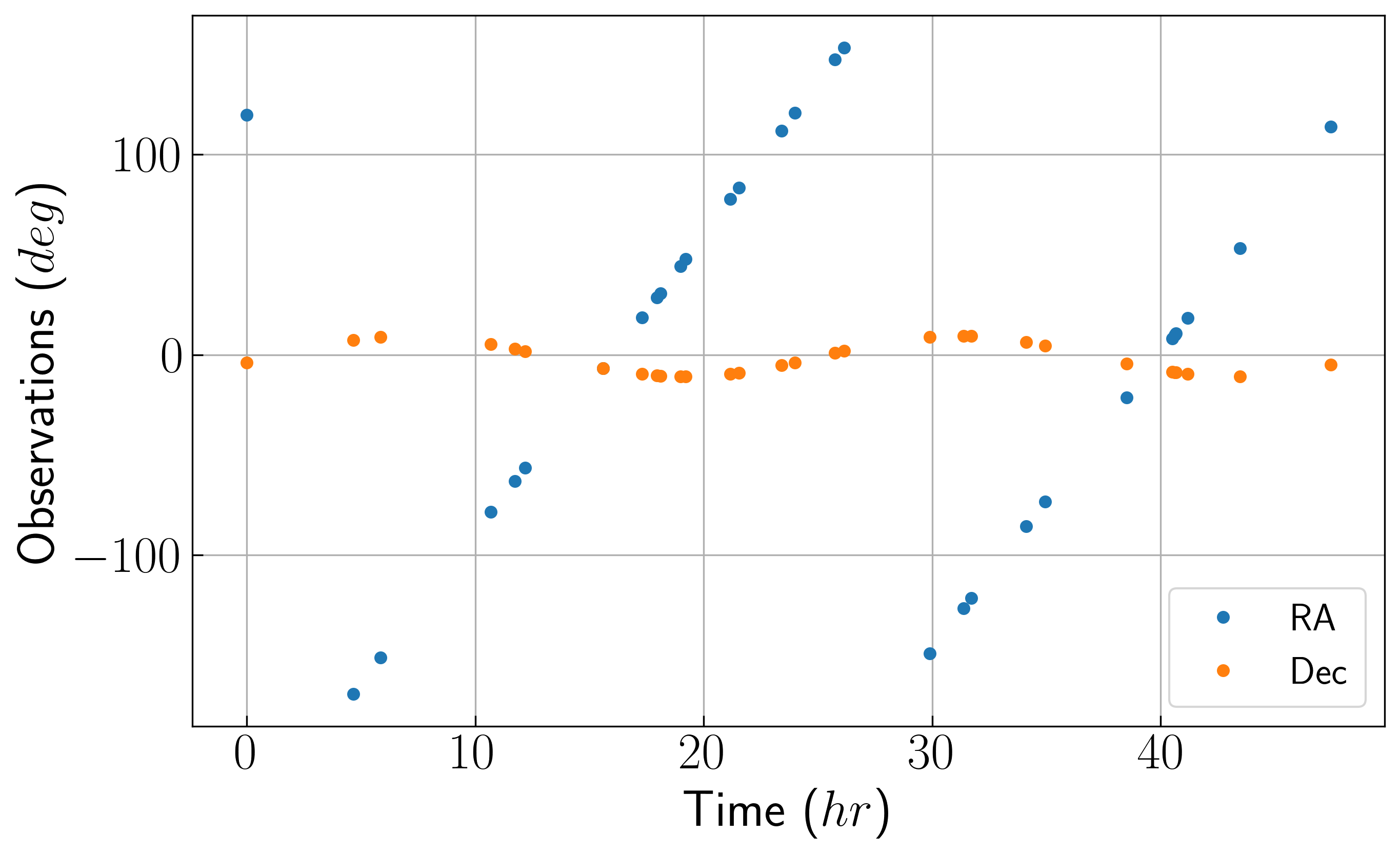}
    \caption{Simulated RA and declination observations of the satellite over the 48 hour time span used for training both the PINN and the physics-only models.}
    \label{fig:raw_obs}
\end{figure}

The simulation creates angles-only observations of a thrusting satellite in GEO as viewed by a ground-based optical telescope. In this demonstration, the data amounted to thirty measurements of right ascension (RA) and declination over the course of two days, randomly sampled in time. Sensor measurement error was added to the observations by randomly sampling a Gaussian distribution with zero mean and 0.5 arcsec standard deviation in both dimensions. The resulting simulated observations in RA and declination are shown in Figure \ref{fig:raw_obs}. Note that daytime outages due to solar exclusion have not been included in the simulated data for simplicity.

\begin{figure}
    \centering
    \includegraphics[width=\linewidth]{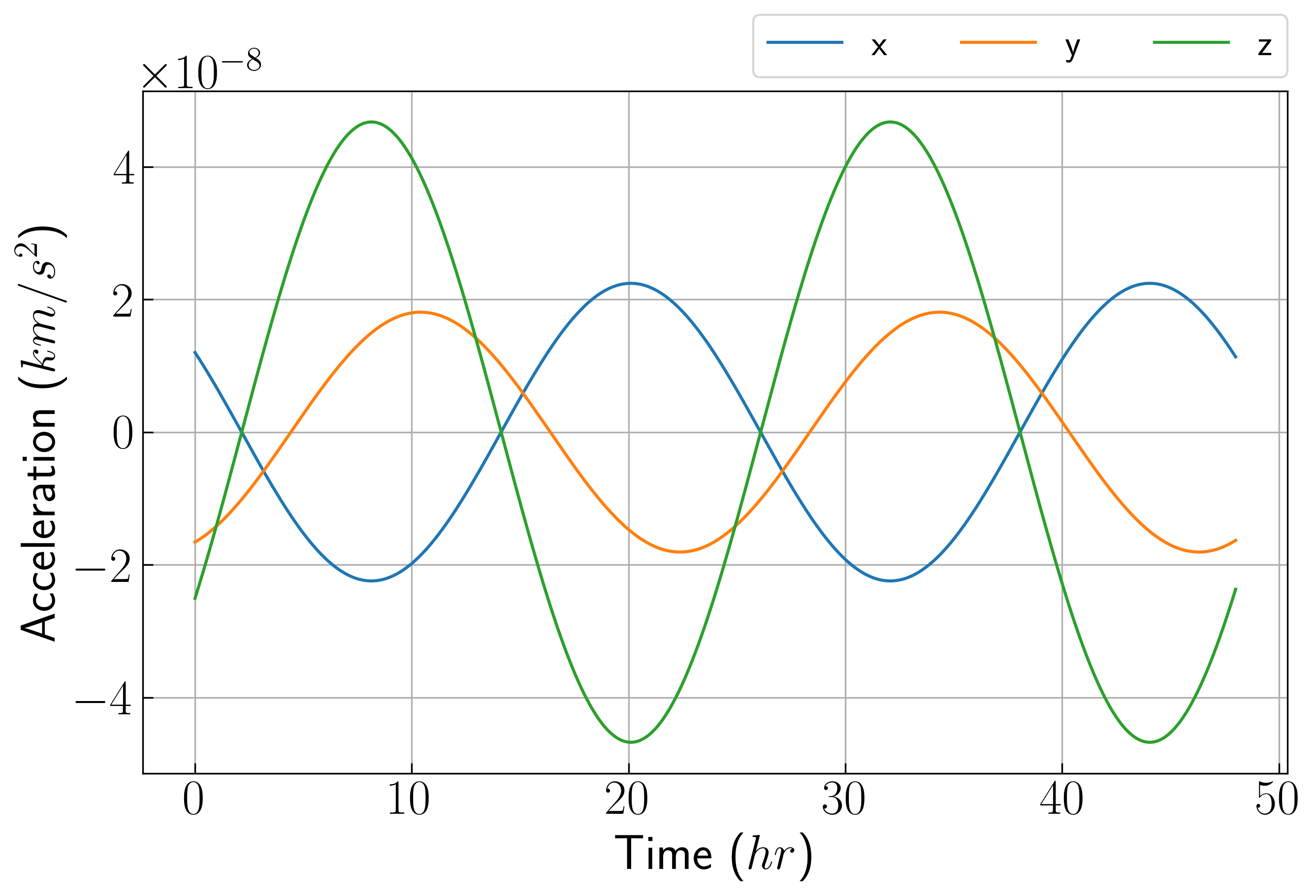}
    \caption{A random periodic acceleration is generated and applied to the satellite in inertial coordinates to simulate an arbitrary thrust profile. The magnitude of the integrated thrust is chosen to lie within the capability of modern electric propulsion systems.}
    \label{fig:periodic_accel_profile}
\end{figure}

The acceleration profile is chosen to be periodic, with periodicity equal to the satellite's orbital period, as shown in Figure \ref{fig:periodic_accel_profile}. This acceleration profile would not be useful for a satellite to achieve a specific change in orbit; however, the goal is to demonstrate that the DNN is capable of learning arbitrary acceleration profiles, and thus the form of the acceleration does not matter. Integrating the magnitude of the thrust vector over the two-day simulation amounts to a total $\Delta$V $\approx$ 10 m/s. 

To understand the feasibility of such a thrust, consider a standard Evolved Expendable Launch Vehicle (EELV) Secondary Payload Adapter (ESPA) satellite with mass of 180 kg \cite{espa_spec}. For such a satellite, a $\Delta$V of 10 m/s with a burn duration of two days would require only 10 mN of thrust, which is well within the capability of most electric propulsion systems. For example, the XIPS engines that are employed on several Boeing 702 class satellites are capable of maximum thrusts of up to 165 mN \cite{goebel2009evaluation}. 

A simple two-body physics model was used along with the acceleration profile described above in order to generate truth data. While this physics model is not accurate enough for most real applications, it is sufficient for a demonstration of the methodology. One of the future goals for this research is to incorporate a higher-fidelity physics model and use the model to fit a state to real observations of a thrusting satellite.

\section{Technical Approach} \label{section:tech_approach}
Ultimately, the goal is to build a model that computes the total acceleration of a satellite in orbit, including both the natural forces and active thrust acting on the satellite. This could be done with a pure physics model, a pure machine learning model, or by combining physics models with machine learning models. This section describes how a physical model--- astrodynamics--- can be coupled with an ML model to describe the total dynamics of a thrusting satellite.

By including physical models in a machine learning framework, the machine learned model need not learn the behavior described by the physics model. For this application, that means we can leverage the astrodynamics models that have been developed and refined over the course of many decades without needing to relearn those governing equations from scratch. The machine learned model need only solve for the deviations from the physical model that best describe the data. As discussed in Section \ref{section:intro}, DNNs are utilized for this work because they can be utilized as parameterized, expressive function approximators. 

To apply knowledge-informed machine learning techniques to the problem of estimating the total dynamics of a thrusting satellite, consider that Equation \ref{eqn:acceleration_components} can be broken apart into two functions:

\begin{equation} \label{eqn:coupled_eom}
    \bm{a} = f(t, \bm{X}) + g(t, \bm{X})
\end{equation}

where $f$ is a physical model incorporating astrodynamics, and $g$ is the instantaneous thrust applied by the satellite, which will be estimated by a DNN.

\begin{figure}
    \centering
    \includegraphics[width=\linewidth]{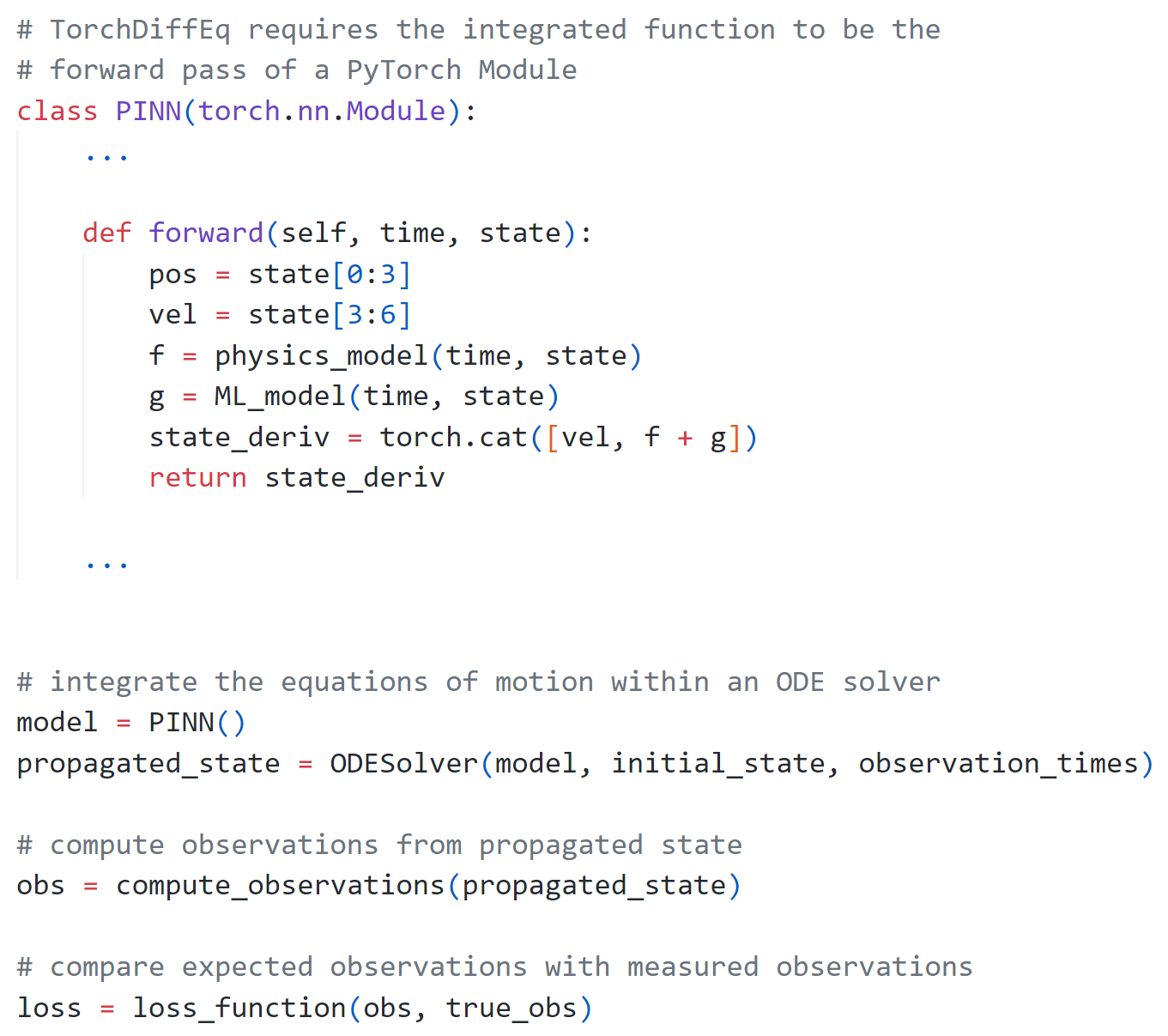}
    \caption{PyTorch- and TorchDiffEq-based pseudocode for the coupled model architecture, orbit propagation, and loss computation. The low-level details of computing observations and the loss function are abstracted.}
    \label{fig:pseudocode}
\end{figure}

As shown in the pseudocode in Figure \ref{fig:pseudocode}, the equations of motion, which combine astrodynamics with the output of a DNN, are integrated using an ODE solver, starting with an initial estimate of the satellite's position and velocity vectors. These initial conditions are also tuned as part of the parameter optimization. The output of the ODE solver is the satellite position and velocity, which can be evaluated at any time within the fit domain. 

\subsection{Model Training} \label{subsec:model_training}
The loss function used for training the coupled model is the mean squared error (MSE) of the observation residuals. The satellite position and velocity vectors are determined by propagating the initial conditions as described above. From those position and velocity vectors, along with the sensor position and velocity at the same times, the expected observations can be computed geometrically. These expected observations are then compared to the true observations and the total loss is computed.

The loss function is back-propagated through the ODE solver in order to determine the gradient with respect to the model parameters. This technique was first demonstrated with the publication of Neural ODEs \cite{chen2018neuralode}. A number of open-source software packages exist for solving neural ODEs, including TorchDiffEq \cite{torchdiffeq} and SciML \cite{rackauckas2019diffeqflux}, \cite{rackauckas2020universal}. The model developed for this effort is implemented in PyTorch \cite{pytorch} and uses TorchDiffEq to back-propagate through the ODE solver.

The initial position and velocity vectors of the satellite are also tuned in addition to the DNN parameters as a part of the model training process. This can be achieved by either a) including the initial conditions as tunable parameters within the model, or b) excluding these parameters from the model and solving for them separately. Both approaches were evaluated as part of this research, and the latter was found to converge more quickly and consistently to the true values of those parameters. Specifically, the initial conditions are estimated using a simple batch least-squares fit to the observation data once every $N$ iterations through the training loop. It was found that fitting the initial conditions every 100 training steps resulted in good performance.

\subsection{DNN Architecture and Hyper-Parameter Tuning}
The acceleration profile to be estimated by the DNN is shown in Figure \ref{fig:periodic_accel_profile}. The accelerations are not particularly complicated and should be well-estimated by a model with relatively few parameters. The results shown in this paper were achieved using a simple Multi-Layer Perceptron (MLP) \cite{MURTAGH1991183} with two hidden layers of 100 nodes each and a hyperbolic tangent activation between each layer. 

As with most machine learning research, the hyper-parameters governing training must be tuned to achieve best performance. For this work, the Adam optimizer \cite{kingma2014adam} is used to tune the ML model parameters and the learning rate (LR) was gradually decreased throughout training using a fixed-step schedule. Table \ref{tab:hyperparams} shows the values of the hyper-parameters that were used for training.

\begin{table}
\renewcommand{\arraystretch}{1.3}
    \centering
    \caption{Hyper-parameters used during training.}
    \begin{tabular}{c|c}
         Parameter & Value \\ \hline
         Training Epochs & 20,000 \\
         Initial Learning Rate & 0.003 \\
         LR Scheduler Step Size & 100 \\
         LR Scheduler Scale Factor & 0.98 \\
         IC Tuning Frequency & 100
    \end{tabular}
    \label{tab:hyperparams}
\end{table}

\begin{figure}
    \centering
    \includegraphics[width=\linewidth]{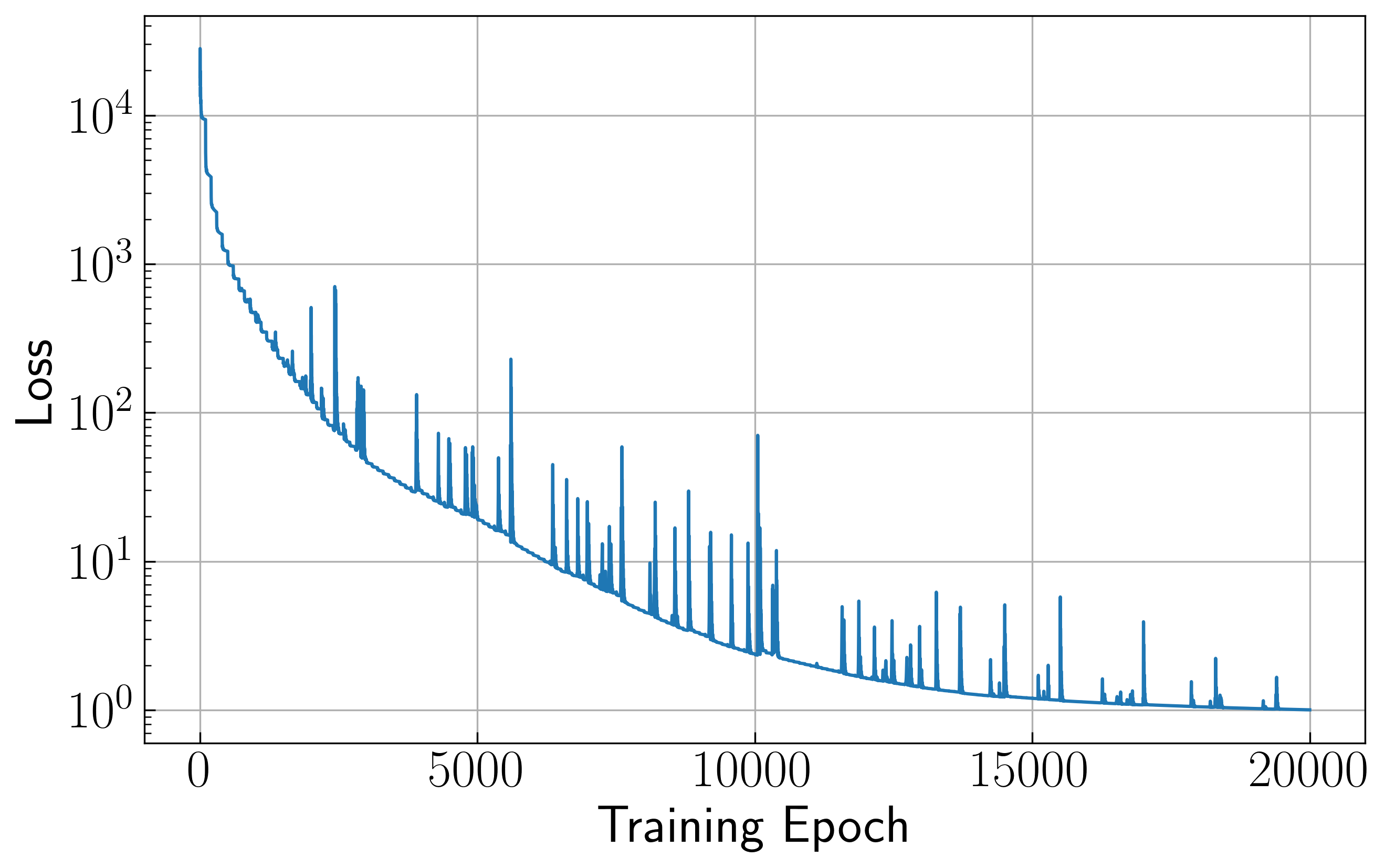}
    \caption{Training loss as a function of training epoch.}
    \label{fig:loss_curve}
\end{figure}

The loss curve obtained during training of the PINN is shown in Figure \ref{fig:loss_curve}. The MSE of the observation residuals decreases by several orders of magnitude before asymptoting to the noise floor of the observations.

The number of training epochs required for this particular data was determined after much experimentation. For this methodology to be used in practice, a more robust training schedule, including early termination, should be implemented to ensure that the model training is sufficient and achieves the expected performance. Additional research is required to evaluate the extent to which the model overfits the data, and how such overfitting can be mitigated. One challenge in this domain is the general lack of data-- observations of slowly maneuvering satellites are sparse as explained in Section \ref{section:intro}-- and thus splitting the data into train and test batches would further reduce the amount of data used for training.

\section{Results} \label{section:results}
Two models were developed and optimized with the loss function described above: 1) a physics-only model that includes two-body equations of motion but does not attempt to model thrust, and 2) the knowledge informed model architecture described in Section \ref{section:tech_approach}. The first model was optimized using simple batch least-squares optimization, while the second was tuned using the Adam optimizer algorithm. These models were then evaluated based on how well they fit the data and how well they were able to predict the true ephemeris in the future, beyond the fit span of the training data.

\subsection{Observation Residuals} \label{subsec:results_residuals}

\begin{figure}
    \centering
    \includegraphics[width=\linewidth]{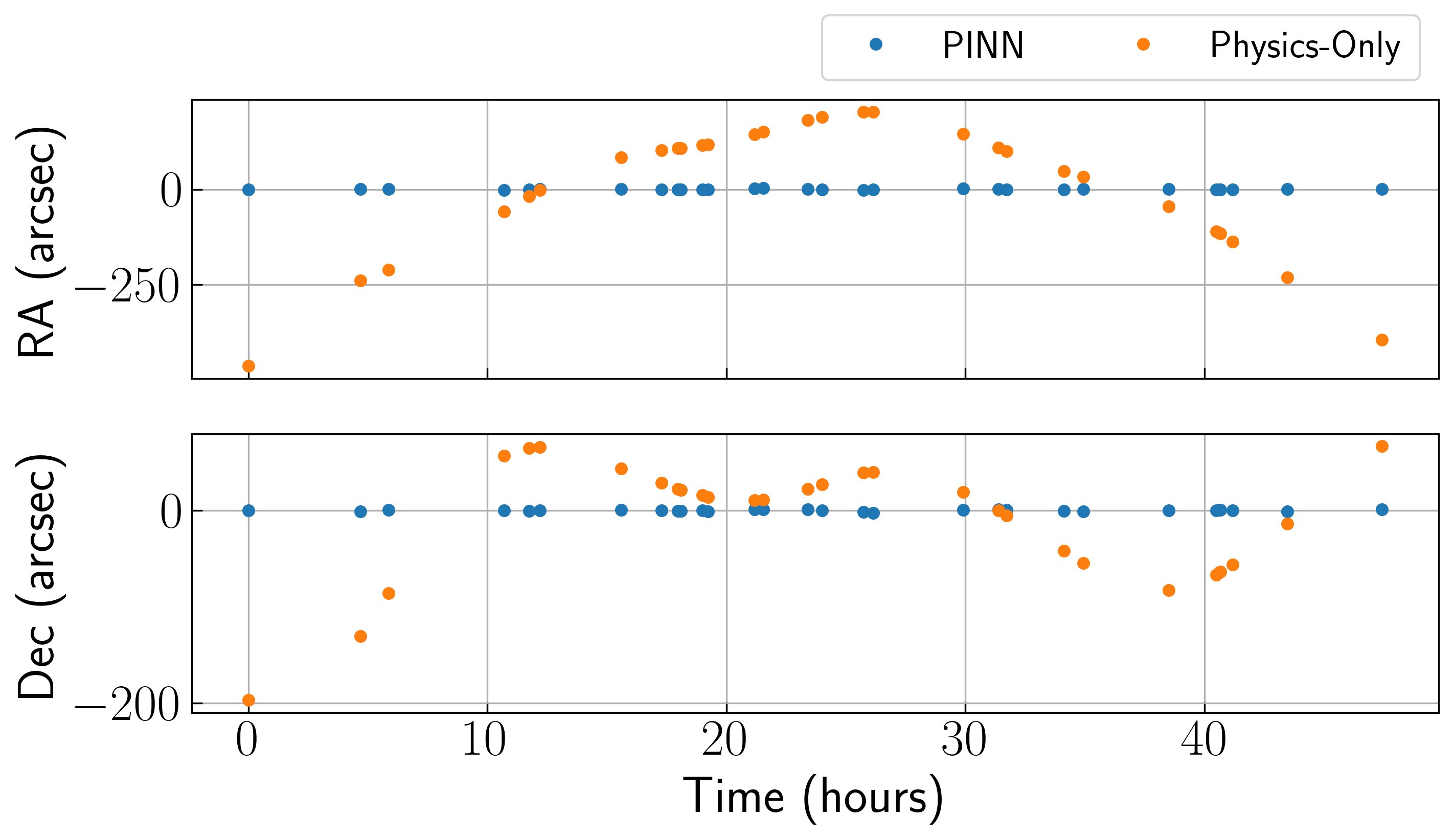}
    \caption{Scatter plot of the RA and declination residuals for both the PINN and physics-only models. The PINN is able to model the arbitrary thrust profile applied, while the physics-only model, which fails to account for thrust, has high residuals over the fit span.}
    \label{fig:residuals_scatter}
\end{figure}

\begin{figure}
    \centering
    \includegraphics[width=\linewidth]{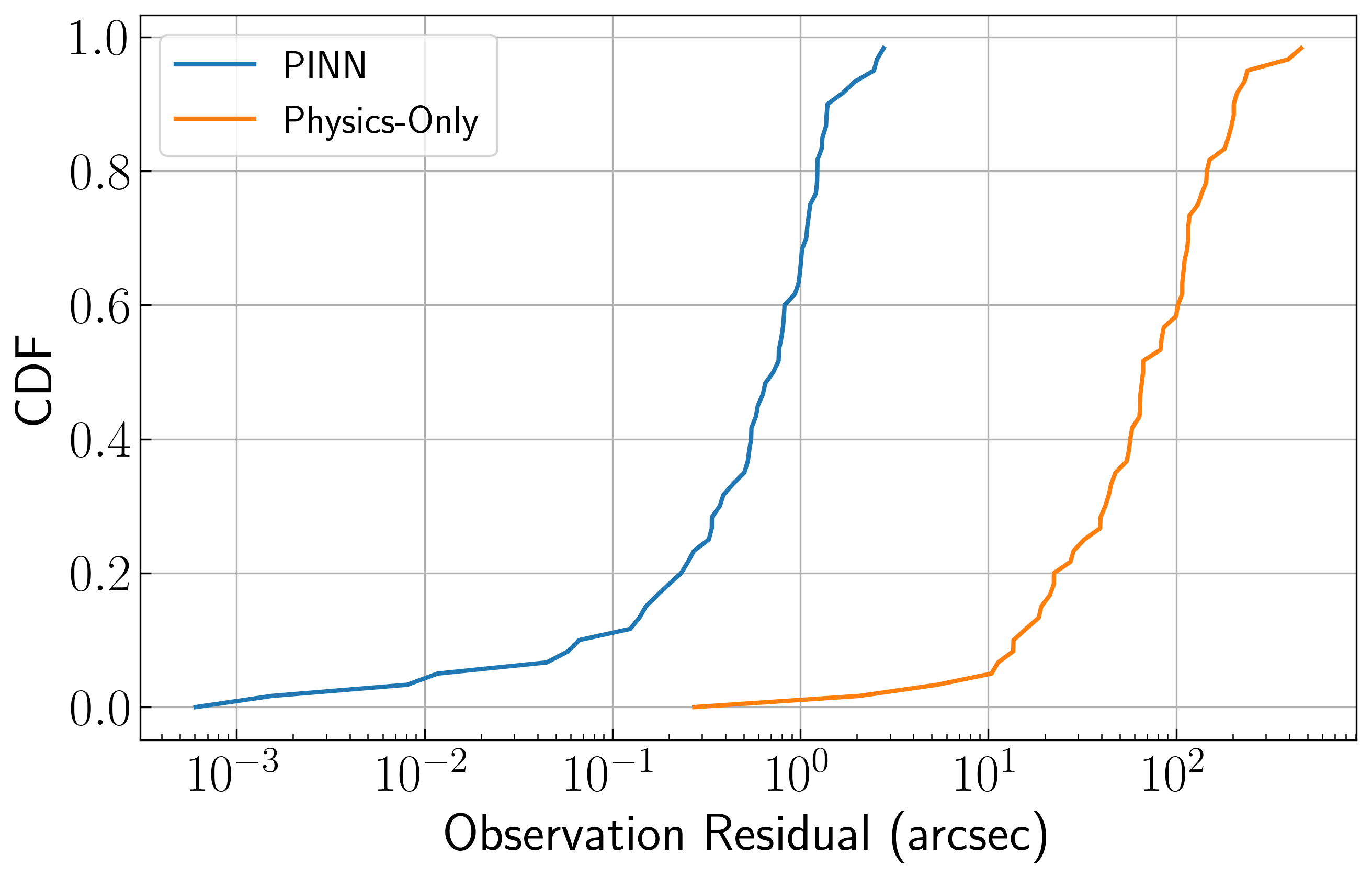}
    \caption{Cumulative distributions of the observation residuals for physics-only and PINN models.}
    \label{fig:residuals_cdf}
\end{figure}

Figure \ref{fig:residuals_scatter} shows a scatter plot of RA and declination observation residuals as a function of time, and Figure \ref{fig:residuals_cdf} combines the residuals for each model into a cumulative distribution. The median observation residual for the physics-informed model is two orders of magnitude smaller than the physics-only model. The physics-informed model has many more degrees of freedom to fit the data, which should result in much lower observation residuals. Additionally, the pure physics model is fundamentally integrating incorrect dynamics by not accounting for the thrust exerted by the satellite. 

The Root Mean Square Error (RMSE) of the PINN was 1.00 arcsec, compared to 123 arcsec for the best-fit physics-only model, an improvement of two orders of magnitude. By simulating observations with random Gaussian noise, characterized by a standard deviation of 0.5 arcsec, the PINN adeptly captures the dynamics of the satellite, effectively extending its explanatory capacity to the approximate noise floor of the observations.

\subsection{Extrapolation Error} \label{subsec:extrapolation}
\begin{figure}
    \centering
    \includegraphics[width=\linewidth]{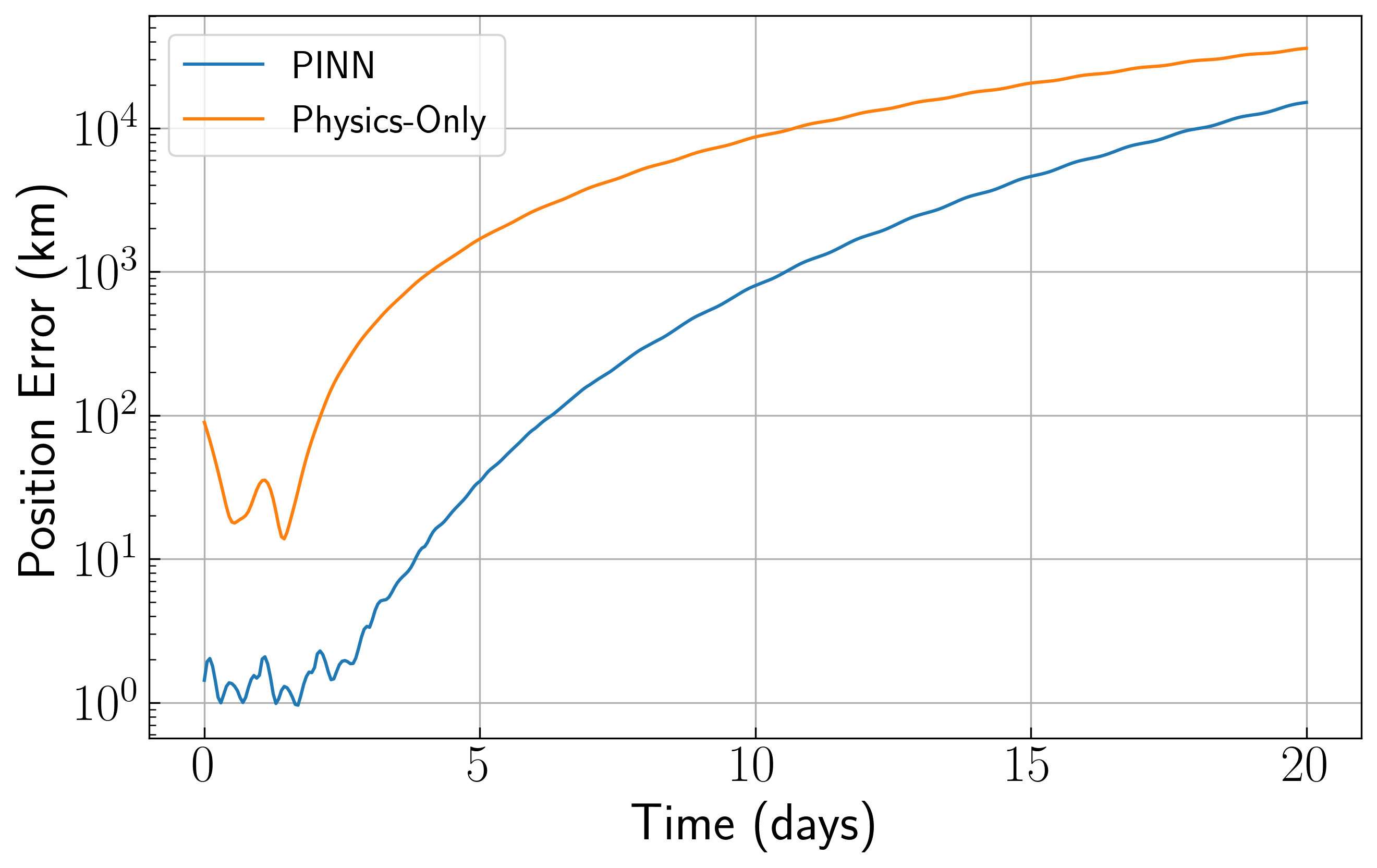}
    \caption{Propagation error beyond the fit span of the observations. While the propagation error grows as a function of time for both models, the PINN model outperforms the physics-only model for the full 20 days of the simulation, sometimes outperforming the physics-only model by as much as two orders of magnitude.}
    \label{fig:extrap_error}
\end{figure}

In order to evaluate how well the models predict the true ephemeris of the satellite in the future, each model was used to propagate the satellite's state for twenty days from the beginning of the simulation and compared with the true propagated position of the satellite. The Euclidian distance between the predicted and true position and velocity was computed. This metric is particularly important for this application because the utility of the model is largely based on how well the model can be used to predict where the satellite will be in the future in order to correlate new observations with the satellite's orbital state. Table \ref{tab:results} gives the propagated position and velocity errors for both one day and five days \textit{after} the end of the two day fit span, while Figure \ref{fig:extrap_error} shows the propagation errors for both models during the full twenty days from the start of the simulation.

\begin{table}
\renewcommand{\arraystretch}{1.3}
    \centering
    \caption{Extrapolation errors of the pure physics and PINN models at 1 and 5 days after the last observation.}
    \begin{tabular}{l|c|c|}
    \cline{2-3}
                                                     & \textbf{Physics-Only} & \textbf{PINN} \\ \hline
    \multicolumn{1}{|l|}{Observation RMSE}           & 123''      & 1.00''                \\ \hline
    \multicolumn{1}{|l|}{Propagation Error - 1 day}  & \begin{tabular}{@{}c@{}} 397 km \\ 29.5 m/s \end{tabular} & \begin{tabular}{@{}c@{}} 3.35 km \\ 0.384 m/s \end{tabular}          \\ \hline
    \multicolumn{1}{|l|}{Propagation Error - 5 day}  & \begin{tabular}{@{}c@{}} 3860 km \\ 285 m/s \end{tabular} & \begin{tabular}{@{}c@{}} 164 km\\ 12.3 m/s \end{tabular}              \\ \hline
    \end{tabular}
    \label{tab:results}
\end{table}

\subsection{DNN Evaluation} \label{subsec:dnn_eval}
One benefit of coupling a physics-based model with a DNN is increased interpretability of the combined model. Evaluating the DNN portion of the combined model results in the best estimate for the satellite thrust as a function of time. For this initial demonstration, the true thrust profile is known because the observation data is simulated, which allows for direct evaluation of how well the model estimated the thrust profile.

\begin{figure}
    \centering
    \includegraphics[width=\linewidth]{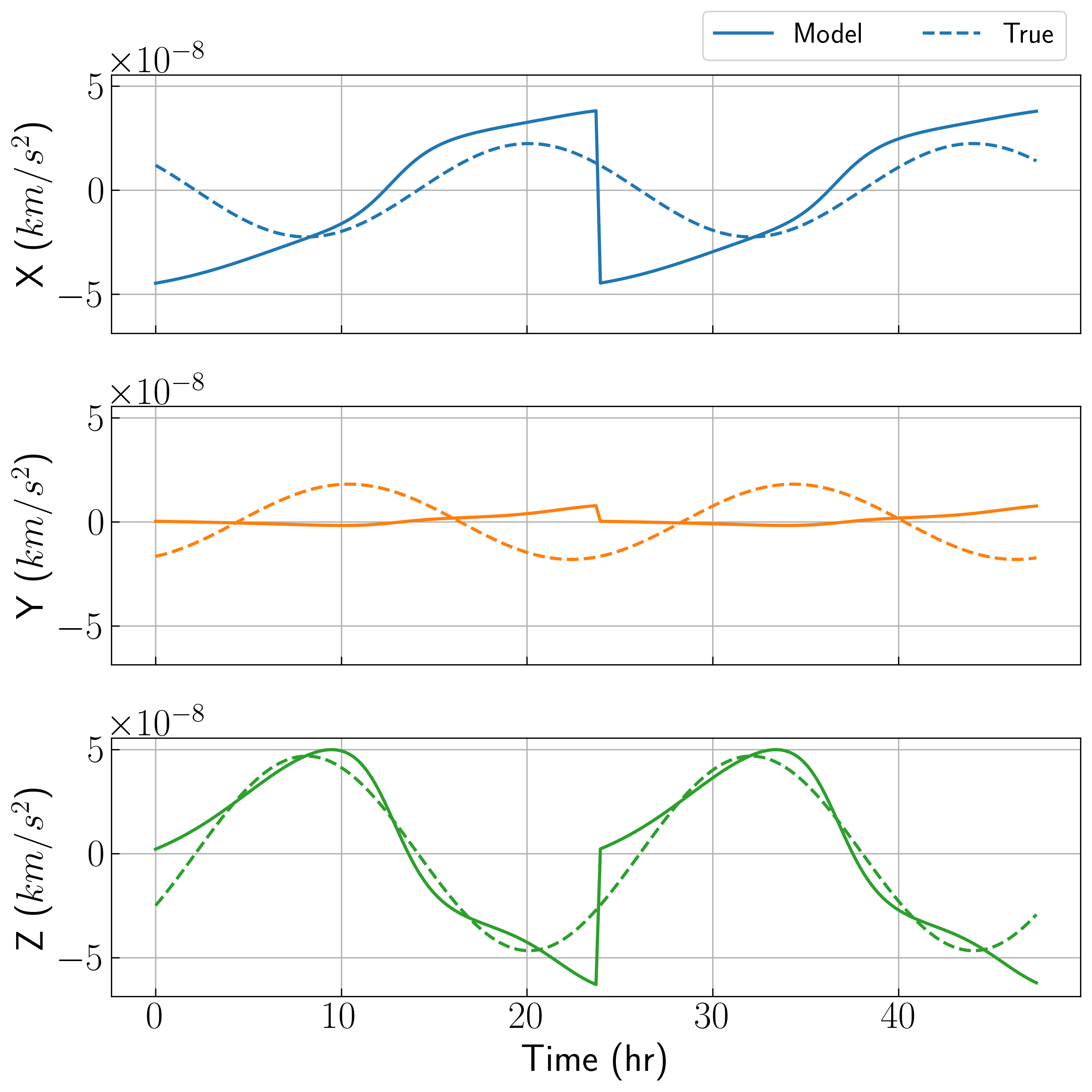}
    \caption{Comparison of the true acceleration profile with the estimated acceleration profile learned by the DNN after training.}
    \label{fig:est_accel}
\end{figure}

Figure \ref{fig:est_accel} shows the components of the estimated thrust as well as the true accelerations used to generate the simulated data. While not perfect, the PINN is clearly capable of learning the general shape and scale of the thrust needed to describe the data. The figure indicates that the DNN estimates the z-component of the thrust profile better than the x- and y-components. It is not inherently clear why the model is better able to fit the z-component than the other directions. One potential reason is that the x- and y- directions are more coupled in the case of a nearly equatorial GEO satellite, and that there is a larger manifold of solutions that explain the observation data equally well. 

It is anticipated that better estimating the individual thrust components would require either more precise observations or longer fit spans, as the ability of the model to accurately estimate the thrust profile depends heavily on the amplitude and duration of the thrust as well as the precision of the observations. The extremely small amplitude thrusts modeled here integrate to a $\Delta$V of only about 10 m/s over a relatively long burn duration of two days; however, even these very small thrusts cause the satellite to deviate from the expected trajectory determined by a pure physics model significantly. 

\section{Summary} \label{section:summary}
An increasing number of satellites are employing low-thrust propulsion systems for their mission needs. This poses a challenge for maintaining satellite catalogs, as orbit propagation tools do not attempt to estimate thrust as part of a satellite's orbital state. This paper demonstrated the use of physics-informed neural networks for learning a satellite's thrust profile along with the typical state vector parameters that best fit a set of angles-only observations of the satellite. The PINN model was shown to produce a better fit to the observations by two orders of magnitude, due to the ability of the DNN to correctly model the satellite thrust which was not captured in the physics-only model. When the trained physics-only and PINN models were extrapolated beyond the last observation, the PINN model again resulted in consistently better performance in estimating the spacecraft's position at future times, sometimes by as much as two orders of magnitude. These results suggest that physics-informed neural networks offer an alternative approach to flexibly model forces acting upon an RSO which are not well-modeled by existing physics-based models and that the inclusion of a DNN component can significantly improve the quality of the resulting orbital state.

\acknowledgments
Research was sponsored by the United States Air Force Research Laboratory and the Department of the Air Force Artificial Intelligence Accelerator and was accomplished under Cooperative Agreement Number FA8750-19-2-1000. The views and conclusions contained in this document are those of the authors and should not be interpreted as representing the official policies, either expressed or implied, of the Department of the Air Force or the U.S. Government. The U.S. Government is authorized to reproduce and distribute reprints for Government purposes notwithstanding any copyright notation herein.

\bibliographystyle{IEEEtran}
\bibliography{refs}

\begin{thebibliography}{10}
\providecommand{\url}[1]{#1}
\csname url@samestyle\endcsname
\providecommand{\newblock}{\relax}
\providecommand{\bibinfo}[2]{#2}
\providecommand{\BIBentrySTDinterwordspacing}{\spaceskip=0pt\relax}
\providecommand{\BIBentryALTinterwordstretchfactor}{4}
\providecommand{\BIBentryALTinterwordspacing}{\spaceskip=\fontdimen2\font plus
\BIBentryALTinterwordstretchfactor\fontdimen3\font minus
  \fontdimen4\font\relax}
\providecommand{\BIBforeignlanguage}[2]{{%
\expandafter\ifx\csname l@#1\endcsname\relax
\typeout{** WARNING: IEEEtran.bst: No hyphenation pattern has been}%
\typeout{** loaded for the language `#1'. Using the pattern for}%
\typeout{** the default language instead.}%
\else
\language=\csname l@#1\endcsname
\fi
#2}}
\providecommand{\BIBdecl}{\relax}
\BIBdecl

\bibitem{18thSDS}
\BIBentryALTinterwordspacing
 [Online]. Available:
  \url{https://www.jtf-spacedefense.mil/About-Us/Fact-Sheets/Display/Article/3155799/18th-space-defense-squadron/}
\BIBentrySTDinterwordspacing

\bibitem{space_policy_directive3}
\BIBentryALTinterwordspacing
U.~S.~W. House, ``Space policy directive-3, national space traffic management
  policy,'' Jun 2018. [Online]. Available:
  \url{https://rosap.ntl.bts.gov/view/dot/60966}
\BIBentrySTDinterwordspacing

\bibitem{shepherd2006space}
L.~C.~G. Shepherd and A.~F.~S. Command, ``Space surveillance network,'' in
  \emph{Shared Space Situational Awareness Conference, Colorado Sprincs, CO},
  2006.

\bibitem{vallado2006revisiting}
D.~Vallado, P.~Crawford, R.~Hujsak, and T.~Kelso, ``Revisiting spacetrack
  report\# 3,'' in \emph{AIAA/AAS Astrodynamics Specialist Conference and
  Exhibit}, 2006, p. 6753.

\bibitem{thomas2016comparison}
D.~Thomas, ``A comparison of geo satellites using chemical and electric
  propulsion,'' 2016.

\bibitem{kiml2023}
L.~von Rueden, S.~Mayer, K.~Beckh, B.~Georgiev, S.~Giesselbach, R.~Heese,
  B.~Kirsch, J.~Pfrommer, A.~Pick, R.~Ramamurthy, M.~Walczak, J.~Garcke,
  C.~Bauckhage, and J.~Schuecker, ``Informed machine learning – a taxonomy
  and survey of integrating prior knowledge into learning systems,'' \emph{IEEE
  Transactions on Knowledge and Data Engineering}, vol.~35, no.~1, pp.
  614--633, 2023.

\bibitem{vallado2001fundamentals}
\BIBentryALTinterwordspacing
D.~Vallado and W.~McClain, \emph{Fundamentals of Astrodynamics and
  Applications}, ser. Fundamentals of Astrodynamics and Applications.\hskip 1em
  plus 0.5em minus 0.4em\relax Microcosm Press, 2001. [Online]. Available:
  \url{https://books.google.com/books?id=OCkGmwEACAAJ}
\BIBentrySTDinterwordspacing

\bibitem{montenbruck_gill}
E.~Gill and O.~Montenbruck, \emph{\BIBforeignlanguage{English}{Satellite
  orbits: Models, methods and applications}}.\hskip 1em plus 0.5em minus
  0.4em\relax Springer, 2013.

\bibitem{espa_spec}
P.~Wegner, J.~Ganley, and J.~Maly, ``Eelv secondary payload adapter (espa):
  providing increased access to space,'' in \emph{2001 IEEE Aerospace
  Conference Proceedings (Cat. No.01TH8542)}, vol.~5, 2001, pp. 2563--2568
  vol.5.

\bibitem{goebel2009evaluation}
D.~M. Goebel, J.~E. Polk, I.~Sandler, I.~G. Mikellides, J.~R. Brophy, W.~G.
  Tighe, and K.-R. Chien, ``Evaluation of 25-cm xips thruster life for deep
  space mission applications,'' in \emph{31st International Electric Propulsion
  Conference (20--24 September 2009}, 2009.

\bibitem{chen2018neuralode}
R.~T.~Q. Chen, Y.~Rubanova, J.~Bettencourt, and D.~Duvenaud, ``Neural ordinary
  differential equations,'' \emph{Advances in Neural Information Processing
  Systems}, 2018.

\bibitem{torchdiffeq}
\BIBentryALTinterwordspacing
R.~T.~Q. Chen, ``torchdiffeq,'' 2018. [Online]. Available:
  \url{https://github.com/rtqichen/torchdiffeq}
\BIBentrySTDinterwordspacing

\bibitem{rackauckas2019diffeqflux}
C.~Rackauckas, M.~Innes, Y.~Ma, J.~Bettencourt, L.~White, and V.~Dixit,
  ``Diffeqflux.jl-a julia library for neural differential equations,''
  \emph{arXiv preprint arXiv:1902.02376}, 2019.

\bibitem{rackauckas2020universal}
C.~Rackauckas, Y.~Ma, J.~Martensen, C.~Warner, K.~Zubov, R.~Supekar,
  D.~Skinner, and A.~Ramadhan, ``Universal differential equations for
  scientific machine learning,'' \emph{arXiv preprint arXiv:2001.04385}, 2020.

\bibitem{pytorch}
A.~Paszke, S.~Gross, F.~Massa, A.~Lerer, J.~Bradbury, G.~Chanan, T.~Killeen,
  Z.~Lin, N.~Gimelshein, L.~Antiga, A.~Desmaison, A.~Köpf, E.~Yang, Z.~DeVito,
  M.~Raison, A.~Tejani, S.~Chilamkurthy, B.~Steiner, L.~Fang, J.~Bai, and
  S.~Chintala, ``Pytorch: An imperative style, high-performance deep learning
  library,'' 2019.

\bibitem{MURTAGH1991183}
F.~Murtagh, ``Multilayer perceptrons for classification and regression,''
  \emph{Neurocomputing}, vol.~2, no.~5, pp. 183--197, 1991.

\bibitem{kingma2014adam}
D.~P. Kingma and J.~Ba, ``Adam: A method for stochastic optimization,''
  \emph{arXiv preprint arXiv:1412.6980}, 2014.

\end{thebibliography}

\thebiography
\begin{biographywithpic}
{Jacob Varey}{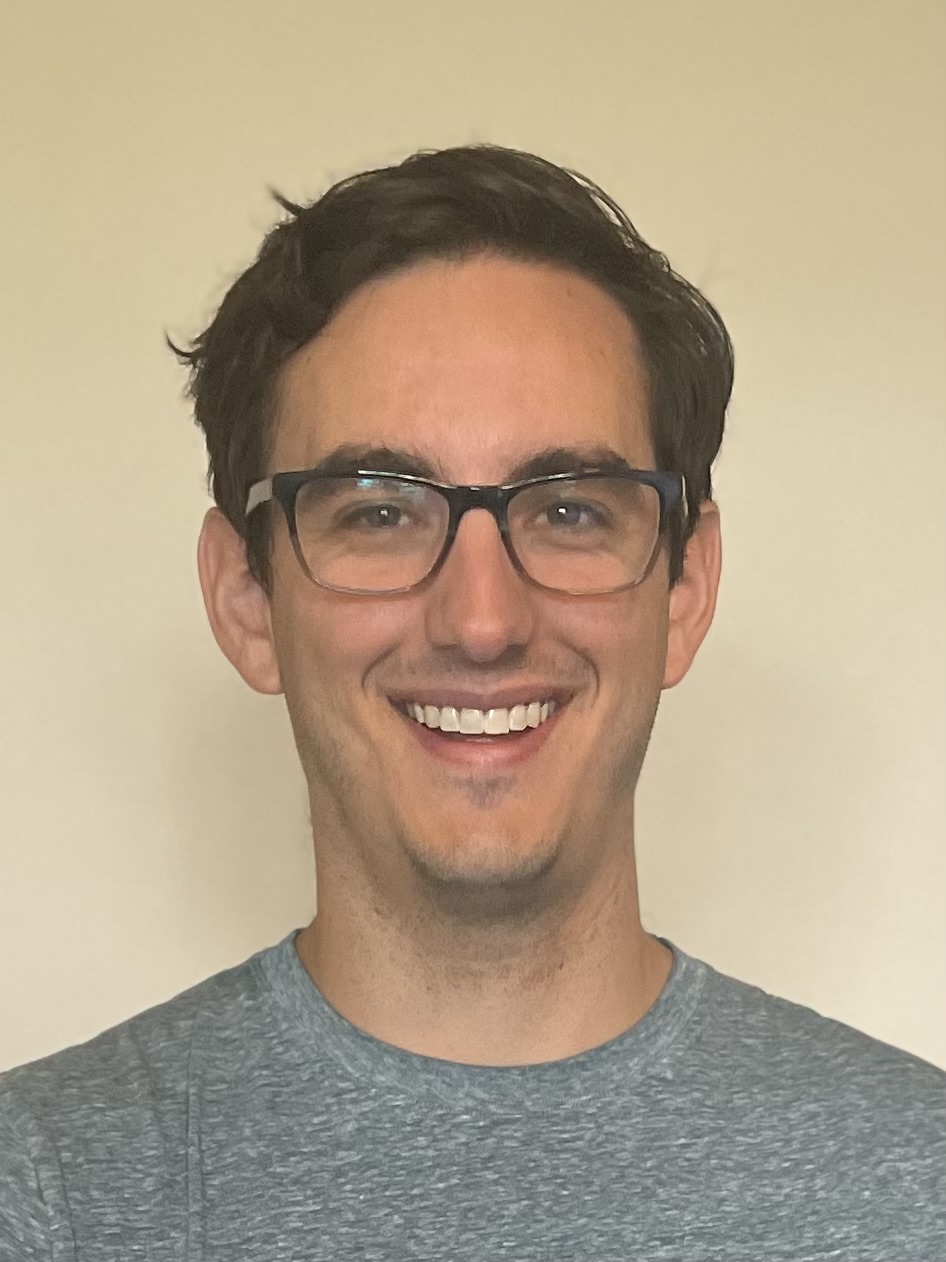}
received his B.S. and M.S. degrees in Aerospace Engineering Sciences with a focus in astrodynamics and satellite navigation from the University of Colorado at Boulder. He is currently a researcher at MIT Lincoln Laboratory in the Artificial Intelligence Technology group. In his eight years with Lincoln Laboratory, his research has focused on remote sensing for space domain awareness, algorithm development for novel approaches to faint signal detection, and deep reinforcement learning in support of decision making in the space domain.
\end{biographywithpic} 

\begin{biographywithpic}
{Jessica D. Ruprecht}{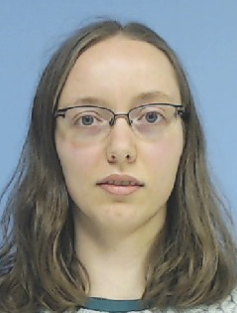}
received a B.S. in Physics and Planetary Science and an M.S. in Planetary Science from the Massachusetts Institute of Technology (MIT). As Technical Staff in the AI Software Architectures and Algorithms group at MIT Lincoln Laboratory, she applies her background in observational astronomy to programs in support of Lincoln Laboratory's space mission. Her current research focuses on applying a variety of artificial intelligence techniques to space problems, including reinforcement learning, computer vision, and natural language processing. 
\end{biographywithpic}

\begin{biographywithpic}
{Michael C. Tierney}{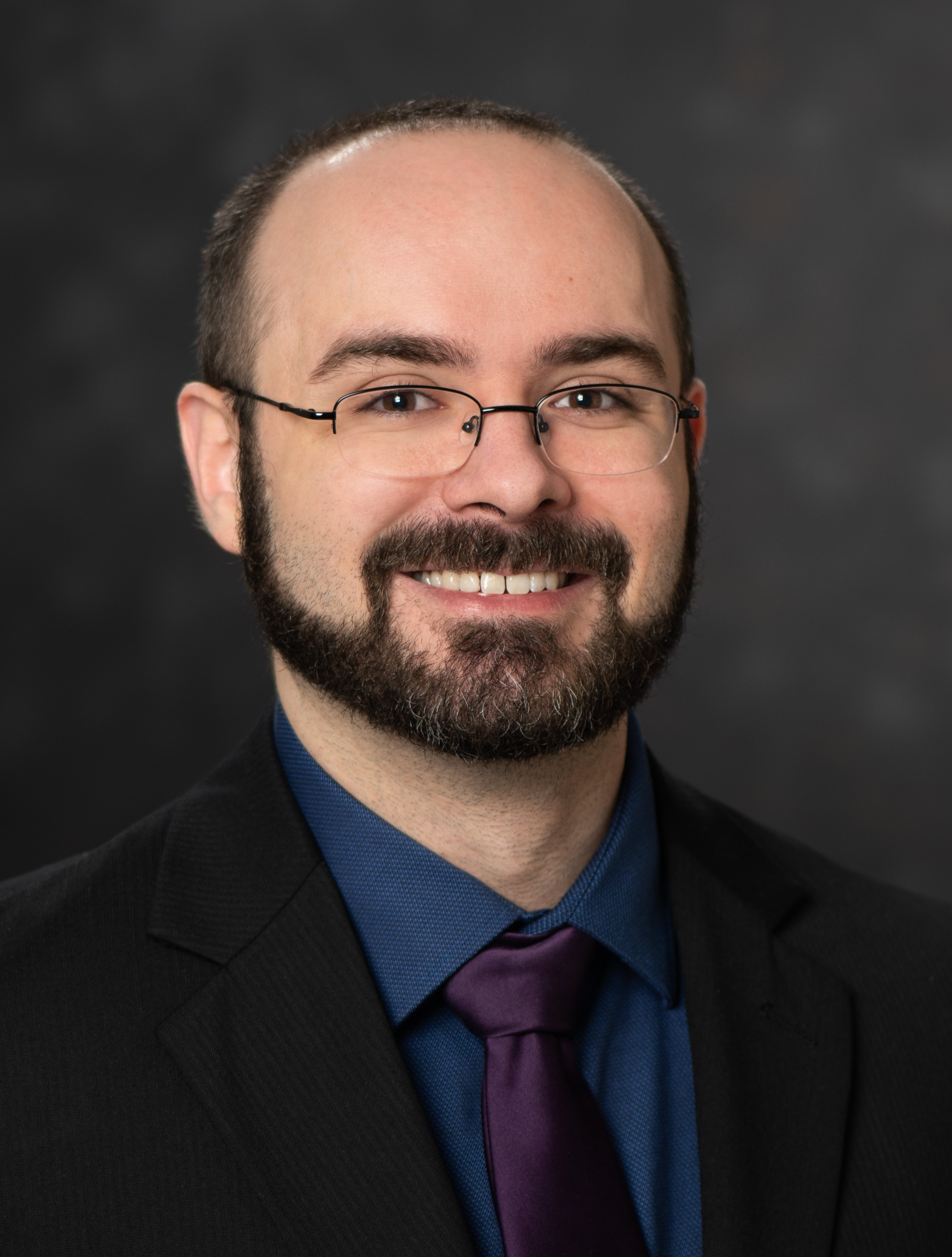}
received his B.S. in Physics from MIT and his M.S. in Astronomy from the University of Hawai`i at M\=anoa. He is currently a member of the MIT Lincoln Laboratory Artificial Intelligence Technology group, where his research interests include building high-fidelity simulators for space domain awareness scenarios, testing the robustness of transformer-based time series classifiers, and building tools to measure the trustworthiness and generalizability of machine learning models.
\end{biographywithpic}

\begin{biographywithpic}
{Ryan M. Sullenberger}{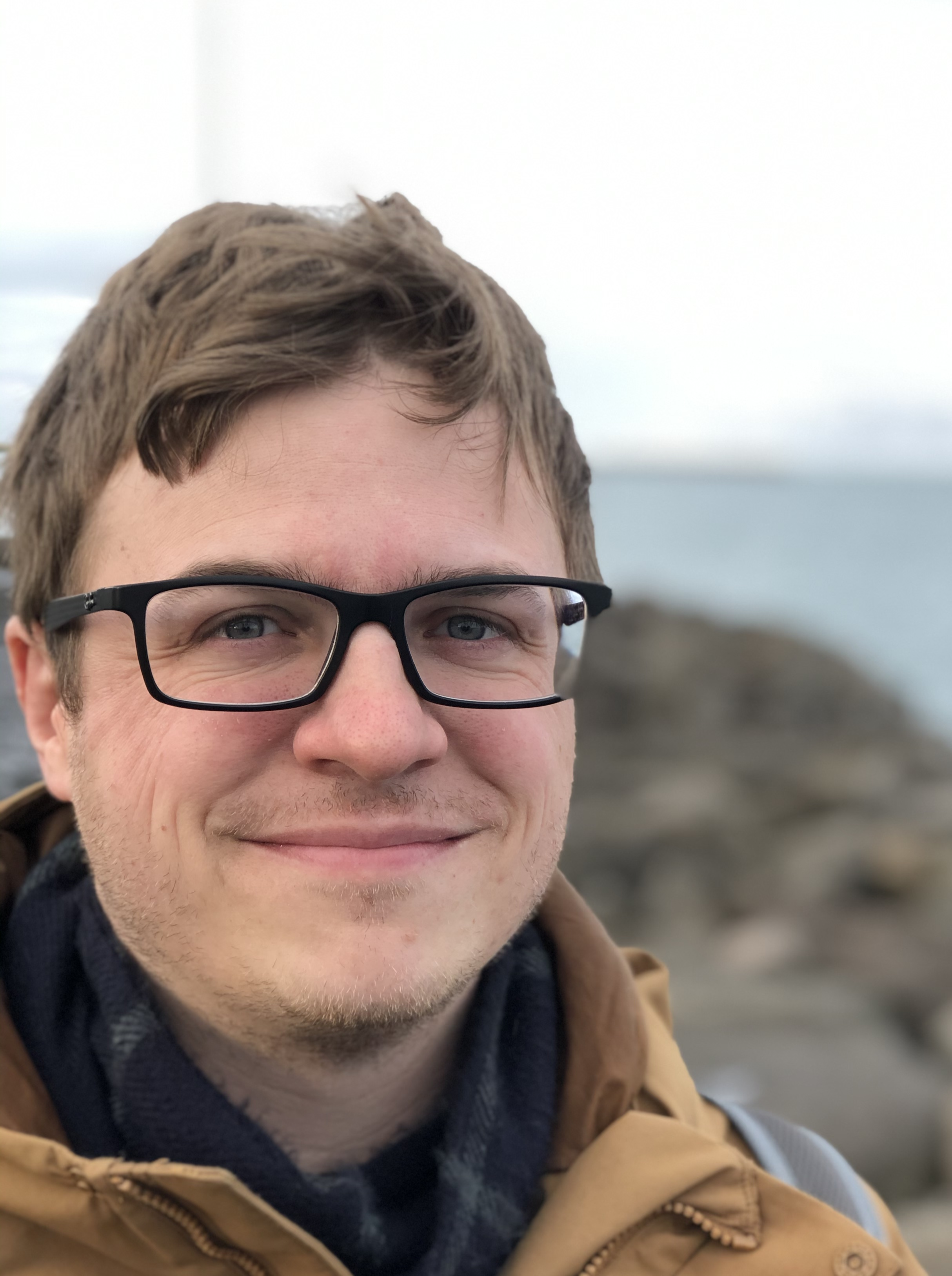}
earned his M.S. in Mechanical and Aerospace Engineering from Princeton University and his B.S. in Mechanical Engineering from Colorado State University. Currently serving as a cross-disciplinary research engineer at MIT Lincoln Laboratory in the Applied Space Systems group, Ryan is involved in the development, design, simulation, analysis and experimental validation of innovative new products and sensors used for detecting and tracking both natural phenomena and human-made space objects. Ryan's areas of expertise and interests include applied physics, remote sensing, rapid prototyping, data science and machine learning.
\end{biographywithpic}

\end{document}